\documentclass{aa}
\usepackage{psfig}
\def\ti{\theta_{\mathrm{I}}}
\def\te{\theta_{\mathrm{E}}}
\def\ts{\theta_{\mathrm{S}}}
\def\ri{r_{\mathrm{I}}}

\def\Scrit{\Sigma_{\mathrm{crit}}}
\def\lya{Ly$\alpha$}     %
\def\nhi{N(\mbox{H\,{\sc i}})}%
\def\Lb{L_{\mathrm B}}   %
\def\kms{~km~s$^{-1}$}   %
\def\cm2{~cm$^{-2}$}     %
\def\zd{z_{\mathrm d}}   %
\def\ze{z_{\mathrm e}}   %
\def\h50{h_{50}^{-1}}    %
\def\hi{H\,{\sc i}}      %
\def\nhi{N(\mbox{H\,{\sc i}})}%
\begin{document}
\thesaurus {03(11.17.1; 11.08.1; 11.09.4)}
\title{Lensing properties of 7 damped \lya\ absorbing galaxy-QSO 
pairs}
\author{Vincent Le Brun\inst{1}, Alain 
Smette\inst{2,3,4}\thanks{Chercheur qualifi\'e du F.N.R.S. (Belgium)}, Jean 
Surdej\inst{4}
\thanks{Directeur de Recherches au F.N.R.S. (Belgium)},
  Jean-Fran\c cois Claeskens\inst{4}\thanks{Charg\'e de Recherches au 
F.N.R.S. 
(Belgium)},}
\institute{Laboratoire d'Astronomie Spatiale du C.N.R.S., B.P. 8, 
F-13376
Marseille, France, Vincent.LeBrun@astrsp-mrs.fr
\and
Laboratory for Astronomy and Solar Physics, NASA-Goddard Space Flight 
Center, 
Code 681, Greenbelt, MD 20771, USA
\and
National Optical Astronomy Observatories, P.O. Box 26732, 950 North
Cherry Avenue, Tucson, AZ 85726-6732, USA
\and
Institut d'Astrophysique et de G\'eophysique, Universit\'e de 
Li\`ege, Avenue 
de Cointe 5, B-4000
Li\`ege, Belgium
}
\offprints{V. Le Brun}
\date {Received ; accepted ?}
\maketitle
\markboth{}{Le Brun et al., Lensing properties of 7 damped \lya\  
absorbing
  galaxy-QSO pairs}
\begin{abstract}
  Le Brun et al. (1997)  presented the first
  identifications of the galaxies giving rise to 7 intermediate
  redshift damped \lya\ (DLA) absorption systems. Here, we study the
  gravitational lensing properties of these foreground galaxies based
  on their observed optical appearance and on the absence of any
  secondary lensed quasar image. We consider the possibility that any
  secondary image be hidden due to extinction by dust, but find it
  unlikely.  We derive upper limits on the amplification factor
  affecting the luminosity of the background quasars; in each case,
  this factor is found to be less than 0.3 mag.  We also obtain upper
  limits on the total mass of the damped Ly-$\alpha$ galaxies, within
  radii equal to the quasar impact parameters.  Mass-to-light ratios
  are found to be consistent with existing estimates based on X-ray
  emission or on motion of dwarf satellites. Although we show that
  lensing is not important in this sample, we note that existing DLA
  surveys used to determine the cosmological density of gas at $z < 1$
  are based on samples of quasars brighter than the ones considered
  here and for which the amplification bias is likely to be 
stronger.  
\keywords{Gravitational lensing -- Galaxies: halos -- Galaxies: ISM -- Quasars:
    absorption lines}
\end{abstract}   
%
 \section{Introduction}
\label{intro}

Redshifted damped hydrogen \lya\ (DLA) lines are sometimes detected in
the spectra of background quasars. They present neutral hydrogen
column densities $\nhi > 2 \times 10^{20}$\cm2\  similar to the value
found at the optical radius of present-day galactic disks (Wolfe et
al. 1986), which suggests that they arise in progenitors of
present-day spiral galaxies.

Because the impact parameters necessary to produce such absorption
lines are likely to be small, the galaxy gravitational potential may
lead to significant gravitational lensing effects, such as
amplification of the apparent luminosity of the background quasar, or
even formation of multiple images. 

In turn, these effects can induce severe biases in the samples of
quasars and absorbers. First, impact parameters are increased because
of lensing (the {\it by-pass\/} effect; see Smette, Claeskens, Surdej 
1997, hereafter, SCS) thus decreasing the
cross-section for DLA absorption. Second, background quasars which
present DLA absorptions in their spectra also have their apparent
luminosity increased because of lensing: hence, intrinsically faint
quasars can enter apparent magnitude limited samples.  As a
consequence, these samples may show an over-representation of quasars
that present DLA absorption systems: this {\it amplification bias\/}
apparently increases the cross-section for damped Ly-$\alpha$
absorption.  These aspects can be evaluated in a statistical sense for
a well-defined magnitude limited sample (cf. 
SCS; Bartelmann \& Loeb 1996). For bright
quasar samples, the amplification bias was found to dominate the
by-pass effect, which leads to predictions that the number density
$d{\cal N}/dz$ of $z \sim 0.5$ DLA systems and the cosmological
density of neutral hydrogen $\Omega_{\mbox{H\,{\sc i}}}$ (since DLA 
absorbers contain most of the \hi\ of the Universe)
 may
be significantly overestimated. Note that extinction by dust 
introduces
other biases competing with the ones just described (Fall \& Pei 
1993, Perna et al. 1997, Bartelmann \& Loeb 1998).



It must be noted however, that existing theoretical estimates are
based on a series of simplifying assumptions, one of which being that
the DLA absorbers are very similar to present-day disk galaxies. As a
matter of fact,  Le Brun et al. (1997) presented the 
first
identifications of damped \lya\  absorbing galaxies at intermediate
redshifts: in a sample of 7 DLA absorbers, 3 turn out to be spiral 
galaxies of
various sizes and luminosities, 2 are compact objects and 2 are
amorphous, low-surface brightness galaxies (an additional compact
object appears to be responsible for a higher-redshift DLA
system). This study showed that the population of galaxies giving
rise to DLA systems is more heterogeneous than previously thought.

Independently, SCS found that lensing effects may be present in the
IUE (Lanzetta et al. 1995) and Rao et al. (1995) Mg\,{\sc ii} surveys.
In addition, in a follow-up of this latter survey, Rao \& Turnshek 
(2000) find a surprisingly high value of $\Omega_{\mbox{H\,{\sc i}}}$ at $ 
z < 1$, as expected if lensing effects are important (SCS).

Here we consider the lensing properties of the absorbing galaxies
detected by Le Brun et al. (1997). These observations were made using 
the WFPC2 camera onboard the Hubble Space Telescope, so that the 
absorbing
galaxy candidates are clearly separated from the quasar images, and
their characteristics (impact parameter, luminosity, morphology)
are fairly well estimated.  

The sample of quasars described here is not complete in any sense
except that it contains all the DLA systems known at the time of
writing the HST proposal (cycle 6). {\it For illustration
  purpose only\/}, we have evaluated the effects of lensing following 
SCS
(exponential model plus Gaussian hole, or E+GH).  These theoretical
estimates are based on the observed magnitude $m_{\mathrm{450}}$ and
emission redshift $\ze$ of the quasar and on the DLA absorber redshift
$\zd$.
Would the DLA sample be representative of  a well-defined, complete
magnitude limited sample, we find that the number density of DLA
systems $d{\cal N}/dz$ would have been overestimated by 18\%.  In
other words, statistically speaking, one of the quasars in this sample
would not have been observed if gravitational lensing was not taking
place.  In addition, the value of $\Omega_{\mbox{H\,{\sc i}}}$ would 
have been
over-estimated by 28\%. We also find that the probability that one
quasar in the sample presents multiple-imaging would be close to 50\%.
The values $H_{\rm 0} = 
100~{\rm km}~{\rm s}^{-1}~{\rm Mpc}^{-1}$ and $(\Omega,\Lambda) = 
(0,0)$
are used throughout this paper.

\section{Lensing properties of the DLA absorbing galaxy/quasar pairs}

The observed characteristics of the seven quasar/galaxy pairs are
presented in Table~\ref{pres}.
\begin{table*}
\caption[]{Characteristics of the seven galaxy/quasar
  pairs: quasar emission redshift $\ze$, apparent magnitude in the
  F450 filter (similar to the B band), DLA line redshift $\zd$, neutral hydrogen
  column density $\nhi$, angular impact parameter 
  $\theta_{\rm I}$, linear projected impact parameter $\ri$, {\it k\/}-corrected
absolute magnitude $M_{\mathrm{B}}$  and its estimated 1$\sigma$ error
 of the absorbing
galaxy (see Boiss\'e et al. 1998 for details). The last column
provides the galaxy morphological type, derived  from the WFPC2 images
as precisely as
possible. 
} 
\label{pres} 
\begin{tabular}{lllllllll}
\hline\noalign{\smallskip}
Quasar name   & $\ze$ & $m_{\mathrm{450}}$ & $\zd$  & $\nhi$ & 
$\ti$    & $r_{\mathrm I}$ & $M_{\mathrm{B}}$  & Type \\
              &       &                    &        
&                          &(\arcsec) &   ($h^{-1}$ kpc)         
&                   &      \\   
EX 0302$-$223 & 1.400 & 16.5               & 1.0095 & 
20.39                    & 1.14     & $6.2\pm0.1$     & 
$-18.1\pm0.1$     & Compact Dwarf\\
PKS 0454+039  & 1.345 & 17.6               & 0.8596 & 
20.69                    & 0.80     & $4.1\pm0.1$     & 
$-18.5\pm0.1$     & Compact Dwarf\\
3C 196        & 0.871 & 18.7               & 0.4370 & 
20.8                     & 1.10     & $4.1\pm0.1$     & 
$-20.0\pm0.05$    & Face-on Giant Spiral\\
Q 1209+107    & 2.191 & 18.6               & 0.6295 & 
20.2                     & 1.62     & $7.3\pm0.1$     & 
$-19.7\pm0.05$    & Edge-on Spiral\\
PKS 1229$-$021& 1.038 & 17.6               & 0.3950 & 
20.75                    & 1.40     & $4.9\pm0.1$     & 
$-16.9\pm0.2$     & Diffuse Dwarf\\
3C 286        & 0.849 & 18.0               & 0.6920 & 
21.19                    & 0.90     & $4.3\pm0.1$     & 
$-18.7\pm0.2$     & Irregular \\
MC 1331+170   & 2.084 & 16.6               & 0.7446 &  
$-$                     & 3.86     & $18.8\pm0.0$    & 
$-20.9\pm0.2$     & Edge-on Bright 
Spiral\\
\noalign{\medskip}\hline
\end{tabular}
\end{table*}
We present three different methods to evaluate the lensing properties
of these seven configurations: {\it (i)\/} a model-independent
formalism; {\it (ii)\/} a formalism which assumes a Singular
Isothermal Sphere (SIS) lens model, only based on the geometry of the
systems (i.e. redshifts of the quasar and the galaxy in each 
association and their angular separation);
{\it (iii)\/} a formalism which assumes a SIS model whose velocity
dispersion is determined from the observed luminosity of the DLA
galaxies and the Tully-Fisher relation. 

The values of the angular impact parameter $\ti$ and absolute
luminosity $\Lb$ of the absorbing galaxy image were obtained by Le
Brun et al. (1997) and Boiss\'e et al. (1998).

No secondary lensed QSO image is detected in any of the systems
presented here at more than 0.3\arcsec\ from the quasar, down to a
limiting magnitude of $m_{450}^{\mathrm{lim}} = 25.5$, i.e. 6.8 to 9 magnitudes
fainter than the main observed QSO image.  The absence of any secondary
image either means that the lensing configuration is indeed not 
capable to
produce multiple imaging, or that the apparent luminosity of the secondary image
is too faint to be detected, possibly due to 
extinction by dust. 

\subsection{(Nearly) model-independent constraints}
\label{sec:independent}
Subramanian \& Cowling (1986) showed that, for a spherical mass
distribution, with surface density decreasing from the center to the
outer parts, as expected for individual galaxy halos, a {\it
  sufficient} and {\it necessary} condition to have multiple images is
that the central value $\Sigma_{\rm o}$ is larger than a 
critical value $\Scrit$ defined by :
\begin{equation}
\label{eq:critical}
\Scrit = \frac{c^2}{4 \pi G} \frac{D_\mathrm{os}}{D_\mathrm{ol}
D_\mathrm{ls}},
\end{equation}
and thus independently of the model chosen for the halo.  In this
relation, $c$ is the speed of light, $G$ the gravitational constant,
$D_\mathrm{os}$, $D_\mathrm{ol}$ and $D_\mathrm{ls}$ the
angular-diameter distances between the observer and the source (QSO),
the observer and the lens, the lens and the source, respectively.
Furthermore, they show that the absence of a secondary lensed QSO
image ensures that $\Sigma_0 < \Sigma_\mathrm{crit}$ still for a
spherical mass distribution. They conjecture
that this result is also valid for centrally peaked elliptical lenses.
We therefore
conclude that the absence of a secondary QSO image at a separation
larger than $0.3 \arcsec$ in our sample ensures that $\Sigma_0 <
\Sigma_\mathrm{crit}$ over such an angular scale.

On the other hand, for a lens with circular symmetry, the mean surface
density within the Einstein radius $\te$ is equal to $\Scrit$ (cf.
Schneider, Falco \& Ehlers 1992); in addition, the location of the
main image is always such that its angular separation from the lensing
galaxy $\ti \ge \te$. In particular, this relation is true even in the
case of multiple images with any secondary image being hidden due to
extinction by dust.  Consequently, we can derive an upper
limit on the projected mass $M_{\mathrm{max}}^{\Sigma}(< \ri)$
enclosed in a disk of radius $r_{\mathrm{I}} = D_\mathrm{ol} ~ \ti$
centered on the galaxy, as
\begin{equation}
\label{eq:m_max}
M_{\mathrm{max}}^{\Sigma}(<\ri) = \pi ~ \ri^2 ~ \Scrit 
\end{equation}
and thus also on the average M/L ratio within $\ri$.

No constraint can be set on the amplification $A$ of the QSO image.

\subsection{SIS lens model with only geometrical constraints}
\label{sec:SIS_geom}

We here assume that the distribution of matter within each galaxy can 
be
described as a singular isothermal sphere (SIS), whose volume mass
density $\rho$ is given as a function of the distance $r$ to the
galaxy center by:
\begin{equation}
\label{eq:rho_r}
\rho(r) = {\sigma_{\rm v}^2\over 2 \pi G}{1\over r^2} ,
\end{equation}
where $\sigma_{\rm v}$ is the 1-dimensional velocity dispersion of 
the SIS.
As a consequence, the total projected mass enclosed in a disk of 
radius $r$ is
\begin{equation}
  \label{eq:sigma_r}
  M(<r) = \frac{\pi ~ \sigma_{\rm v}^2}{G}~ r.
\end{equation}

We then compute the angular Einstein radius for each quasar/galaxy 
configuration, by 
\begin{equation}
\label{eq:theta_e}
\theta_{\mathrm E} = 4 \pi \frac{\sigma_{\rm v}^2}{c^2} 
\frac{D_\mathrm{ls}}
{D_\mathrm{os}}.
\end{equation}
In the case of an SIS, the observed 
images are located at:
 \begin{equation}
   \label{eq:SIS}
   \ti =  \ts \pm \te,
 \end{equation}
where $\ts$ is the true (unobserved) position of the source.

Multiple (double) imaging  only occurs for $\ti < 2 \te$, in which
case the separation between the two images is $\Delta \theta = 2 \te$,
and their amplification is $A_{\mathrm I} = (\te/\ts) \pm 1$. 
If $\ti > 2 \te$ only one image is formed and its amplification is
$A_{\mathrm I } = (\te/\ts) + 1$.

Inversely, even in the case of an isothermal sphere with a 
core-radius,
Narayan \& Schneider (1990) showed that the condition $\ti \ge 2\te$
is sufficient to avoid the formation of multiple images. 
Using only the geometry of the system, we can set the following
constraint on the mass of the DLA galaxy. We successively consider 
the cases of
single and double image systems : 

\subsubsection{Single image system}
If only one image is present, $\ti > 2 ~ \te$. Using Eq. 
\ref{eq:sigma_r} and 
inverting Eq. \ref{eq:theta_e} imply that:
\begin{equation}
  \label{eq:SIS_geom_1image}
  M_{\mathrm{SIS},1}^\mathrm{geom}(< \ri ) < \frac{c^2}{8~G}
  ~ \frac{D_\mathrm{os}}{D_\mathrm{ls} D_\mathrm{ol}} ~ \ri^2 
= \frac{M_{\mathrm{max}}^{\Sigma}(<\ri)}{2}.
\end{equation}
The only constraint on the amplification affecting the observed image
is $ A < 2$.

\subsubsection{Double image system}
Suppose now that two images are actually produced by the lens. In 
order to explain the observations, we have to assume that only the 
brightest image is detected, while the faintest one is affected by 
dust extinction by an amount $E_{\mathrm{B}}$ so that its apparent 
magnitude is fainter than the limiting magnitude of the corresponding 
WFPC2 observations.

The angular separation  between the observed image and the center of 
the lensing galaxy $\ti$ is such that $\te \leq \ti \leq 2\te$. 
Consequently, the total projected mass located within a disk of 
radius $\ri$ is constrained by :
\begin{equation}
  \label{eq:SIS_geom_2images}
  \frac{M_{\rm max}^{\Sigma}(< \ri)}{2} \le M_\mathrm{SIS,2}^\mathrm{ 
geom}(< \ri) \le M_\mathrm{max}^{\Sigma}(< \ri).
\end{equation}

Let's define $\Delta m_{\mathrm{obs}}$ as the difference between the 
limiting magnitude of the corresponding WFPC2 frame and the 
magnitude of the main observed QSO image : $\Delta m_{\mathrm{obs}} = 
m_{450}^\mathrm{lim} - m_{450}(\mathrm{QSO})$. We then have the 
following relations between the amplifications of the two images I and 
B of the quasar :

\begin{equation}
    \label{eq:ampli}
    \frac{A_\mathrm{B}}{A_\mathrm{I}}  =
\frac{\theta_\mathrm{B}}{\theta_\mathrm{I}}
 \le
 10^{-0.4 [\Delta m_\mathrm{obs}  - (E_\mathrm{B} - E_\mathrm{I}) ]},
\end{equation}
where $E_\mathrm{I}$ and  $E_\mathrm{B}$ are the extinction affecting 
the main and secondary images, respectively.

Since the secondary image can be very close to the center of the 
deflecting galaxy, it might be extremely extinguished, and thus 
undectable on our data. 

 
\subsubsection{Summary}
Only using geometrical quantities with a SIS lens model does not
greatly improve the limits based on a model-independent formalism: the
value $M_{\mathrm{max}}^{\Sigma}(<\ri)$ is within a factor of two the
upper limit of the mass of a SIS lens only constrained by the angular
separation between the lensing galaxy and the quasar, independently of
the fact that multiple images exist or not.

\subsection{SIS lens model constrained by the Tully-Fisher relation}
\label{sec:SIS_TF}

In this section, we consider one additional piece of information 
brought by
the HST observations: the luminosity of the galaxies considered to be
responsible for the DLA absorption.

If we assume that the 1-dimensional velocity dispersion of the SIS
$\sigma_\mathrm{v}$
is equal to the velocity dispersion of matter in the galaxy
$\sigma_\mathrm{v}^\mathrm{TF}$, 
which is related to
the maximal value $v_{\rm max}$ of the rotational
velocity of a galaxy by $\sigma_{\rm v}^{\rm TF} = v_{\rm
  max}/\sqrt{2}$, we can indeed use the Tully-Fisher relation (Tully
\& Fisher 1977) to derive its value from the galaxy luminosity:
\begin{equation}
\label{tf}
\sigma_{\rm v}^{\rm TF} = \sigma_{\rm v}^{\star,\mathrm{TF}} 
     \left( {L\over L_{\rm B}^\star}\right)^{1/\alpha_{\mathrm{TF}}} ,
\end{equation}
where $\sigma_{\rm v}^{\star,\mathrm{TF}}  =144^{+8}_{-13}$~\kms , 
$M_\star(B^0_{\mathrm{T}}) = -19.9^{+0.2}_{-0.4} +
5\log h$ ($h$ is the Hubble constant in units of 100\kms~Mpc$^{-1}$),
and $\alpha_{\mathrm{TF}} = 2.6\pm0.2$; these values are in 
accordance with
the work by Fukugita \& Turner (1991) and SCS.  This relation is
derived from local galaxies, but most recent studies (see Bershady
1996 for a review) indicate that there is no evolution of the
parameters in this relation at intermediate redshifts.  Then, even if
the value of $M_\star$ is known to evolve with redshift (see e.g Lilly
et al. 1995), this only indicates an evolution of the density of
galaxies of a given luminosity, but not of the relation between the
luminosity and the velocity dispersion of a given galaxy.

We can then set an upper limit to the projected mass enclosed in a
disk of radius $\ri$ centered on the galaxy, as Eq. \ref{eq:sigma_r} 
becomes
\begin{equation}
M_{\mathrm{SIS}}^{\mathrm{TF}}(<\ri) = 
\frac{\pi~(\sigma_{\rm v}^{\rm TF})^2}{G} ~ \ri.
\end{equation}

We can also  estimate the amplification factor $A_\mathrm{I}^\mathrm{TF}$
of the quasar apparent luminosity due to gravitational 
lensing by
the galaxy responsible for the DLA absorption, as $\te^{\rm TF}$ is 
derived by
inserting $\sigma_{\rm v}^{\rm TF}$ in Eq. \ref{eq:theta_e}:
\begin{equation}
  \label{eq:A_TF}
  A_{\mathrm I}^{\rm TF} = \frac{1}{1 - 4 \pi \frac{(\sigma_{\rm
        v}^{\rm TF})^2}{c^2} \frac{D_{\rm ls} ~ D_{\rm ol}}{D_{\rm os} ~ \ri}}.
\end{equation}


The values of $M_\mathrm{SIS}^\mathrm{TF}(<r_\mathrm{I})$ and 
$A_\mathrm{I}^\mathrm{TF}$ derived for each of the QSO--galaxy
association  are presented in Table 2. 
As expected, the inferred mass-to-light ratios $M_\mathrm{
SIS}^\mathrm{TF}(< r_{\mathrm I})/L_{\mathrm B}^\mathrm{tot}$, also
listed in Table 2, are close to the mass-to-light ratios for spiral
and elliptical galaxies at the present epoch estimated by various, classical
methods (see Bahcall et al 1997 for a review).  This
result, discussed further in Section 3.3, ensures us that the
values obtained for the amplification factor $A_{\mathrm I}^\mathrm{
TF}$ are probably good estimates.

\begin{table*}[!ht]
\caption[]{\label{results} Lensing properties of the seven
  quasar/galaxy pairs. 
Symbols are defined in the text. The $^{TF}$ upper-script indicates 
that
the corresponding quantity has been evaluated with the Tully-Fisher
relation.  Calculations 
have been done using $H_{\mathrm 0} = 100$ km/s/Mpc, $\Omega=0$, 
$\Lambda
  = 0$, and no correction for extinction.}
\begin{tabular}{ccccccccccc}
\hline\noalign{\smallskip}
\multicolumn{1}{c}{$L/L_B^\star$} & 
\multicolumn{1}{c}{$\sigma_{\rm v}^{\rm TF}$}& 
\multicolumn{1}{c}{$\theta_{\mathrm{E}}^{\rm TF}$}& 
\multicolumn{1}{c}{$\theta_{\mathrm{I}}$}      & 
\multicolumn{1}{c}{$\Sigma_{\mathrm{crit}}$}& 
\multicolumn{1}{c}{$M_{\mathrm{max}}^{\Sigma}(< \ri )$}       &
\multicolumn{1}{c}{$M_{\mathrm{SIS}}^{\rm TF}(< \ri )$}       &
\multicolumn{1}{c}{$\frac{M_{\mathrm{max}}^{\Sigma}(< \ri 
)}{L_{\mathrm{B}}^{\mathrm{tot}}}$} & 
\multicolumn{1}{c}{$\frac{M_{\mathrm{SIS}}^{\rm TF}(< \ri 
)}{L_{\mathrm{B}}^{\mathrm{tot}}}$} & 
\multicolumn{1}{c}{$A_{\mathrm{I}}^{\rm TF}$} \\
                              &
\multicolumn{1}{c}{(\kms)}    &
\multicolumn{1}{c}{(\arcsec)} & 
\multicolumn{1}{c}{(\arcsec)} & 
\multicolumn{1}{c}{(g cm$^{-2}$)}          & 
\multicolumn{1}{c}{($10^{10} ~ M_{\mathrm{\sun}})$}                 &
\multicolumn{1}{c}{($10^{10} ~ M_{\mathrm{\sun}})$}                 &
\multicolumn{1}{c}{($M_{\mathrm{\sun}}/L_{\mathrm{\sun}}$)}         & 
\multicolumn{1}{c}{($M_{\mathrm{\sun}}/L_{\mathrm{\sun}}$)}   &\\
\hline\noalign{\smallskip} 
\multicolumn{10}{c}{EX 0302-223}\\
0.19$^{+0.07}_{-0.04}$ &  76$^{+11}_{ -6}$ & 0.03$^{+0.01}_{+0.01}$ & 
1.14 & 1.715 &  99$\pm$ 2 &  2.6$^{ +0.8}_{ -0.4}$ & 393.5 & 10.4$^{ 
+3.0}_{ -1.7}$ & 1.03$^{+0.01}_{+0.01}$ \\
\multicolumn{10}{c}{}\\

\multicolumn{10}{c}{PKS 
0454+039}\\                                          
0.27$^{+0.10}_{-0.06}$ &  88$^{+13}_{ -7}$ & 0.05$^{+0.02}_{-0.01}$ & 
0.80 & 1.340 &  34$\pm$ 1 &  2.3$^{ +0.7}_{ -0.4}$ & 93.5 & 6.3$^{ 
+1.8}_{ -1.0}$ & 1.07$^{+0.02}_{-0.01}$ \\
\multicolumn{10}{c}{}\\

\multicolumn{10}{c}{3C 
196}\\                                                
1.09$^{+0.41}_{-0.21}$ & 149$^{+21}_{-11}$ & 0.26$^{+0.07}_{-0.04}$ & 
1.10 & 1.128 &  29$\pm$ 1 &  6.6$^{ +1.9}_{ -1.0}$ &  19.5 &  4.6$^{ 
+1.3}_{ -0.7}$ & 1.30$^{+0.05}_{-0.03}$ \\
\multicolumn{10}{c}{}\\

\multicolumn{10}{c}{Q 
1209+107}\\                                            
0.83$^{+0.31}_{-0.16}$ & 134$^{+19}_{-10}$ & 0.26$^{+0.07}_{-0.04}$ & 
1.62 & 0.741 &  59$\pm$ 1 &  9.5$^{ +2.7}_{ -1.4}$ & 54 & 8.7$^{ 
+2.5}_{ -1.2}$ & 1.19$^{+0.03}_{-0.01}$ \\
\multicolumn{10}{c}{}\\

\multicolumn{10}{c}{PKS 
1229-021}\\                                          
0.06$^{+0.03}_{-0.02}$ &  50$^{ +8}_{ -5}$ & 0.04$^{+0.01}_{-0.01}$ & 
1.40 & 0.952 &  35$\pm$ 1 &  0.9$^{ +0.3}_{ -0.2}$ & 416 & 10.6$^{ 
+3.4}_{ -2.1}$ & 1.03$^{+0.01}_{-0.01}$ \\
\multicolumn{10}{c}{}\\

\multicolumn{10}{c}{3C 
286}\\                                                
0.33$^{+0.14}_{-0.09}$ &  94$^{+15}_{ -9}$ & 0.03$^{+0.01}_{-0.01}$ & 
0.90 & 2.628 &  71$\pm$ 2 &  2.7$^{ +0.9}_{ -0.5}$ & 162 & 6.2$^{ 
+1.9}_{ -1.2}$ & 1.04$^{+0.01}_{-0.01}$ \\
\multicolumn{10}{c}{}\\

\multicolumn{10}{c}{MC 
1331+170}\\                                           
2.50$^{+1.03}_{-0.65}$ & 205$^{+32}_{-21}$ & 0.53$^{+0.17}_{-0.11}$ & 
3.86 & 0.794 & 420$\pm$ 2 & 57.3$^{+18.2}_{-11.5}$ & 126 & 
17.3$^{+5.5}_{ -3.5}$ & 1.16$^{+0.03}_{-0.02}$ \\
\noalign{\medskip}\hline
\end{tabular}
\end{table*}

\section{Discussion}
\label{discussion}

\subsection{Absence of a secondary image}
\label{sec:secondary}

As can be seen in Table 2, application of the Tully-Fisher relation
to the absolute luminosity of the DLA galaxies implies that the impact
parameter of the line-of-sight to the QSO never falls within twice
the value of the galaxy Einstein radius. Hence, we do not a priori
expect any secondary lensed QSO images, which is confirmed by the
HST/WFPC2 observations. 

However, the total luminosity of the galaxies might be underestimated, 
because of extinction due    to the
presence of diffuse dust in the galaxies themselves (self-extinction), 
especially in the rest-frame $B$ band in 
which our galaxies are observed. 
Since the Tully--Fisher relation given
above in Eq. \ref{tf} was determined from a sample of local galaxies whose
$B$ magnitudes were corrected for
self-extinction, even for inclined systems, we are led to use 
a global extinction. 

In particular, a significant self-extinction 
correction might be applied to a  DLA galaxy absolute $B$ magnitude
if it presents a large inclination.
Thus, from the 
observed values, we  can calculate the 
self-extinction which is necessary to give a real Einstein radius 
large enough to lead to $\ti \leq 2\te$ and the formation of  a double image.
For four absorbing galaxies, this value of the self-extinction is larger than 3 
magnitudes, which we consider to be implausible, given that such 
values have not been observed in samples dedicated to the study of 
self-extinction in galaxies (Xu et al. 1997).

For the three remaining galaxies (toward 3C 196, Q 
1209+107 and MC 1331+170), the self-extinction necessary to lead to a 
value of $\te$ compatible with multiple imaging is smaller, ranging 
from 1 magnitude for 3C 196 to 1.6 and 1.8 magnitude for the last two. 
The last two values are comparable to the highest ones detected in the 
sample of Xu et al. (1997). 
On the other hand, we consider unlikely that the galaxy responsible
for the DLA in the spectrum of 3C 196 is affected by 1 magnitude of 
self-extinction: the galaxy is seen face-on and its
redshift is $z_\mathrm{d} = 0.4370$, so that the F702W band is
actually centered at 4900\AA\, in the galaxy rest-frame.  Furthermore,
the galaxy extent as determined in the F450W image, which roughly
corresponds to near UV in the galaxy rest frame, is nearly equal to
the one seen in the F702W filter. The self-extinction correction is
thus probably small in this object.


In summary, the self-extinction required to produce multiple imaging
in these 3 systems are unlikely to be significant. However, they are not 
unrealistic.  
Let's assume,
therefore, that multiple imaging is taking place. Consequently, the absence of
detected secondary images allows us to provide some constraints on the
extinction in these galaxies along the particular lines-of-sight to
the QSO images.

Indeed, if multiple imaging is taking place for 3C 196, Q1209+107
and MC1331+170 and if the SIS model is an
 adequate representation of the matter distribution within the lens,
 then the observations could reveal two images with a magnitude
 difference $\Delta m_{\mathrm{BI}}$ smaller than $\Delta m_{\mathrm{
obs},i} - \Delta E$, where $i = 1,2,3$ represents each of the three 
quasars. From Eq. 9, such a situation occurs if

 \begin{equation}
    \theta_{\mathrm S} \le f_\theta ~ \theta_{\mathrm E},
 \end{equation}
 where
 \begin{equation}
    f_\theta =
 \frac{
 1 - 10^{-0.4 ~\Delta m_{\mathrm{BI}} }
 }
 {
 1 + 10^{-0.4 ~\Delta m_{\mathrm{BI}} }
 }.
 \end{equation}

 The probability $P_i(\Delta E; \Delta m_{\mathrm{obs},i})$
 of \textit{not} detecting a secondary image in the  system $i$
 is thus  given by
 \begin{equation}
         P_i = 1 - \frac{\pi ~ \theta_{\mathrm S}^2}{\pi ~ \theta_{\mathrm
E}^2}
              = 1 - (f_\theta)^2.
 \end{equation}

 This last relation assumes that the sources are uniformly distributed
behind the lenses: we neglect the amplification bias, which tends to
select systems with small $\theta_{\mathrm S}$ and, consequently, to
strengthen the following conclusion.

\begin{figure}
\centerline{\psfig{figure=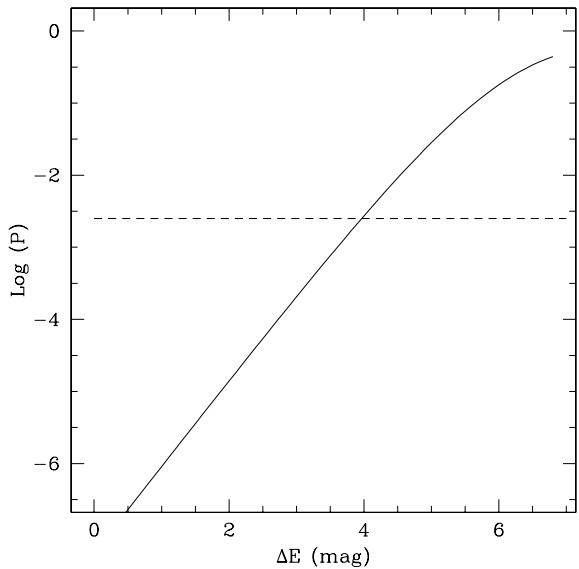,height=8cm,clip=t,angle=0}}
\caption[]{Probability $P$ of not detecting any secondary image for the three 
selected quasars (see text) as a function of the differential 
extinction $\Delta E$
toward the three sight-lines. The horizontal dashed line represents the 
3$\sigma$ (or 0.25\%) level, which we choose as a threshold}
\label{ext} 
\end{figure}

 The probability $P(\Delta E; \Delta m_{\mathrm{obs},1}, \Delta
 m_{\mathrm{obs},2}, \Delta m_{\mathrm{obs},3})$ of not detecting a
 secondary image in any of the 3 systems is therefore
 \begin{equation}
     P = \prod_{i=1}^{3} P_i,
 \end{equation}
 which is represented as a solid line in Fig.~\ref{ext}.

We can see
that the non-detection hypothesis, i.e. the observations, is ruled out
with a confidence level larger than 3 sigma if the differential
extinction is smaller than 3.9 magnitudes on each sightline. But an
extinction larger than 3.9 mag is only expected in very dense clouds,
whose covering factor is very small.

Furthermore, Table 2 shows that the limits on the mass-to-light ratios
inferred from lensing and geometrical constraints alone are usually an
order of magnitude larger than the ones derived from an application of
the Tully-Fisher relation.  If a secondary image was hidden due to
extinction, it would imply that the mean $M/L$ ratio of galaxies is
much higher than previously thought.  

For all these reasons, we maintain our conclusion that the observed 
configurations of
DLA absorbers are not likely to give rise to multiple images.

\subsection{Biases due to gravitational lensing}
\label{sec:biases}

A quasar located at an impact parameter slightly larger than the
Einstein radius can still have its apparent luminosity increased by a
factor A, due to gravitational lensing amplification.  Column~10 of
Table~\ref{results} presents this factor estimated for each of our
systems: as one can see, it is always smaller than 0.3 mag., with an
average value of 0.12 mag. The values for a  $(\Omega,\Lambda) =
(0.3,0.7) $ Universe are only slightly larger (by 3 to 6\%). 

As we now have an estimated value for the amplification, instead of
estimating statistical lensing effects, we can actually compute
lensing effects for each QSO individually.  We used the approach of
Narayan (1989) to evaluate the excess of quasars close to foreground
galaxies.  This method takes into account both the amplification bias
and the by-pass effect. We point out, however, that it aims at
estimating the excess number of quasars in the vicinity of galaxies,
not the excess of quasars in the vicinity of galaxies giving rise to
DLA systems in quasar absorption spectra. However, we prefer this
method due to uncertainties in the determination of the inclination of
the galaxies and other observational variables, and because it is good
enough for our purpose.

The mean excess of quasars close to the DLA galaxies is found to be
equal to a factor $\simeq 1.14$ for a $(\Omega,\Lambda) = (0,0)$ 
Universe.
Therefore, if these quasars were drawn out of a magnitude limited
sample, 14\% of the quasars that present a DLA line in their spectrum
would have been observed because of gravitational lensing (this
calculation should be considered as a mere exercise, as the sample is
not complete). This value is close to the 18\% obtained by the method
described in SCS (cf. Sect.~1).

On the other hand, if these quasars were drawn out of a {\it 
volume\/} limited
sample, the magnification bias would be irrelevant: only the by-pass
effect would be acting.  In this case, the fact that the
observed impact parameters $\theta_{\rm I}$ are larger than
$\theta_{\rm S}$ leads to an {\it underestimate\/} of 
$\Omega_{\mbox{H\,{\sc i}}}$ by about 6\%, based on the E+GH model of 
SCS.

\subsection{Comparison of the $M/L$ ratio constraints with other 
methods}
\label{sec:comparison}

\begin{figure}
\centerline{\psfig{figure=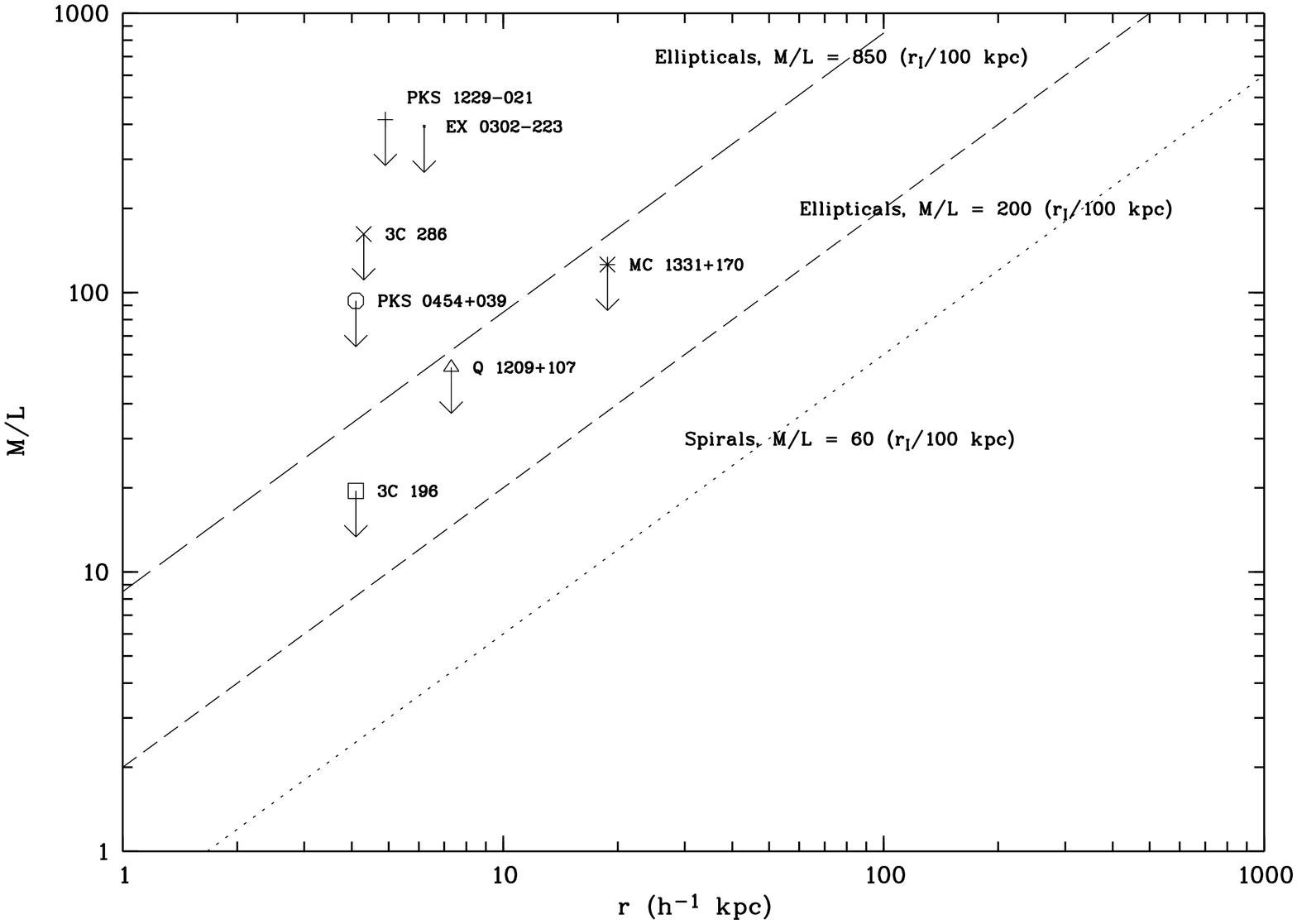,height=8cm,clip=t,angle=-90}}
\caption[]{Upper limits on the mass-to-light ratio $M_{\mathrm{max}}^{\Sigma}(< \ri 
)/L_{\mathrm{B}}^{\mathrm{tot}}$ derived
    from the lensing properties of damped \lya\  absorbing galaxies. 
By comparison, we also show the results from classical methods (cf. 
Bahcall et al. 1995). They indicate that the mass-to-light ratio as a 
function of radius can be represented by power laws, plotted here as
    dotted and dashed lines  for the spiral and elliptical 
galaxies,
    respectively.}
\label{M_sur_L} 
\end{figure}

We have plotted in Fig.~\ref{M_sur_L} the upper
limits to the model-independant mass-to-light ratios $M_{\mathrm{max}}^{\Sigma}(< \ri 
)/L_{\mathrm{B}}^{\mathrm{tot}}$ we derive for the seven galaxies.
The dotted and dashed lines show the power laws representing the
overall mass-to-light ratio in spiral and elliptical galaxies,
respectively, as estimated from Bahcall et al. (1995): this review
presents recent results obtained on the mass-to-light ratio using
the classical methods (\hi\  emission, X-ray emission, motion of 
dwarf satellites ...). 

As can be seen, the strongest limits set by gravitational lensing,
which are obtained for the three spirals of the sample, are an order
of magnitude above the values obtained from other methods. Thus,
although they are compatible with other estimations, these
observations do not allow to efficiently constrain the value of the
hidden mass in the galaxies responsible for the DLAs.

\subsection{Future work}
\label{sec:future}

We note that the published surveys to determine the
cosmological density of neutral hydrogen $\Omega_{\mbox{H\,{\sc i}}}$ 
at $0 < z <
1$ have been carried on using samples of quasars that are generally
brighter than all the quasars on which the present study is based:
only 15\% of the quasars in the IUE survey (Lanzetta et al. 1995) have
a $B$ magnitude fainter than the quasars presented here; half of the
Rao et al. (1995) quasars (followed up by Rao \& Turnshek (2000)) are
brighter than the brightest quasar in our sample, and none are
fainter than the faintest quasar in this same sample (some quasars are
actually used both in this paper and in Rao et al.'s). These surveys
may thus be more strongly affected by the magnification bias than the
sample presented here; consequently, the results of this paper should
not be interpreted as meaning that lensing effects have negligible
effects on surveys of DLAs at $0 < z < 1$.

In order to have a better constraint on the lensing effects in DLA
surveys,  we have carried a HST-NICMOS survey of 13 bright quasars
whose spectra present a DLA system at low redshift, including the 
sample presented in this paper. 

\begin{acknowledgements}
    Our research was supported in part by PRODEX (Gravitational lens studies
with HST), by contract P4/05 "P=F4le d'Attraction Interuniversitaire" (OSTC,
Belgium), by contract 1994-99 of "Action de Recherches Concert=E9es"
(Communaut=E9 Fran=E7aise, Belgium) and by the "Fonds National de la Recherc=
he
Scientifique" (Belgium).
\end{acknowledgements}

\end{document}